\documentclass[sigconf,authorversion,screen]{acmart}
\usepackage{algorithm2e}
\usepackage{listings}
\usepackage[british]{babel}

\AtBeginDocument{%
  \providecommand\BibTeX{{%
    \normalfont B\kern-0.5em{\scshape i\kern-0.25em b}\kern-0.8em\TeX}}}

\setcopyright{acmcopyright}
\copyrightyear{2019}
\acmYear{2019}

\acmConference[EOOLT 2019]{EOOLT 2019: 9th International Workshop on Equation-Based Object-Oriented Modeling Languages and Tools}{November 04--05, 2019}{Berlin, DE}
\acmBooktitle{EOOLT 2019: 9th International Workshop on Equation-Based Object-Oriented Modeling Languages and Tools, November 04--05, 2019, Berlin, DE}
\acmISBN{978-1-4503-7713-3}

\begin{document}

\title{ModelicaGym: Applying Reinforcement Learning to Modelica Models}

\author{Oleh Lukianykhin}
\authornote{Both authors contributed equally to this research.}
\email{lukianykhin@ucu.edu.ua}
\author{Tetiana Bogodorova}
\email{bogodorova@ucu.edu.ua}
\affiliation{%
  \institution{The Machine Learning Lab,}
  \institution{Ukrainian Catholic University}
  \city{Lviv}
  \country{Ukraine}
  \postcode{79026}
  }


\begin{abstract}
  This paper presents $ModelicaGym$ toolbox that was developed to employ Reinforcement Learning (RL) for solving optimization and control tasks in Modelica models.  The developed tool allows connecting models using Functional Mock-up Interface (FMI) to OpenAI Gym toolkit in order to exploit Modelica equation-based modeling and co-simulation together with RL algorithms as a functionality of the tools correspondingly. Thus, $ModelicaGym$ facilitates fast and convenient development of RL algorithms and their comparison when solving optimal control problem for Modelica dynamic models. Inheritance structure of $ModelicaGym$ toolbox's classes and the implemented methods are discussed in details. The toolbox functionality validation is performed on Cart-Pole balancing problem. This includes physical system model description and its integration using the toolbox, experiments on selection and influence of the model parameters (i.e. force magnitude, Cart-pole mass ratio, reward ratio, and simulation time step) on the learning process of Q-learning algorithm supported with the discussion of the simulation results. 
\end{abstract}



\begin{teaserfigure}
  \includegraphics[width=\textwidth]{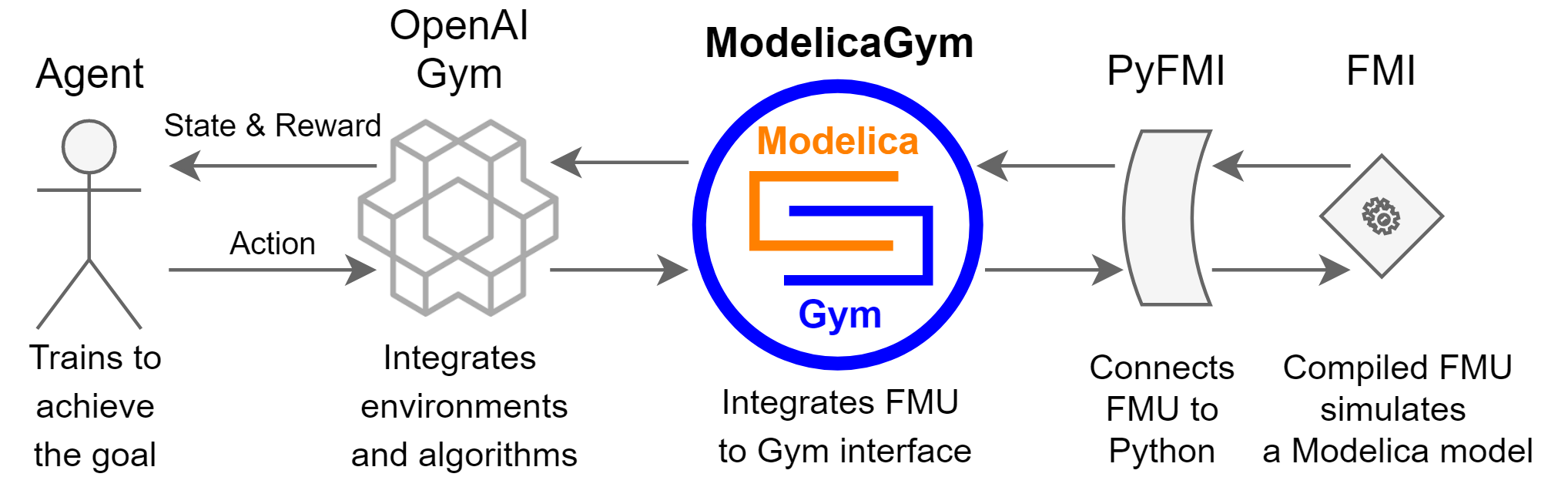}
  \caption{A high-level overview of a considered pipeline and place of the presented toolbox in it.}
  \Description{Overview of how implemented toolbox helps to use FMI for RL.}
  \label{fig:teaser}
\end{teaserfigure}

\maketitle

\section{Introduction}

\subsection{Motivation}
In the era of big data and cheap computational resources, advancement in machine learning algorithms is naturally raised. These algorithms are developed to solve complex issues, such as predictive data analysis, data mining, mathematical optimization and control by computers. 

The control design is arguably the most common engineering application \cite{bogodorova2015bayesian}, \cite{turitsyn2011options}, \cite{AchControlElLoads}. This type of problems can be solved applying learning from interaction between controller (agent) and a system (environment). This type of learning is known as reinforcement learning \cite{sutton2018reinforcement}. Reinforcement learning algorithms are good in solving complex optimal control problems \cite{moriyama2018reinforcement}, \cite{smottahedi-rl}, \cite{hallen2018comminution}.

Moriyama et al. \cite{moriyama2018reinforcement} achieved 22\% improvement compared to a model-based control of the data centre cooling model. The model was created with EnergyPlus and simulated with FMUs \cite{FMUref}.

Mottahedi \cite{smottahedi-rl} applied Deep Reinforcement Learning to learn optimal energy control for a building equipped with battery storage and photovoltaics. The detailed building model was simulated using an FMU.

Proximal Policy Optimization was successfully applied to optimizing grinding comminution process under certain conditions in \cite{hallen2018comminution}. Calibrated plant simulation was using an FMU.

However, while emphasizing stages of the successful RL application in the research and development process, these works focus on single model integration. On the other hand, the authors of \cite{moriyama2018reinforcement}, \cite{smottahedi-rl}, \cite{hallen2018comminution} did not aim to develop a generalized tool that offers convenient options for the model integration using FMU. Perhaps the reason is that the corresponding implementation is not straightforward. It requires writing a significant amount of code, that describes the generalized logic that is common for all environments. However, the benefit of such implementation is clearly in avoiding boilerplate code instead creating a modular and scalable open source tool which this paper focused on.

OpenAI Gym \cite{brockman2016openai} is a toolkit for implementation and testing of reinforcement learning algorithms on a set of environments. It introduces common Application Programming Interface (API) for interaction with the RL environment. Consequently, a custom RL agent can be easily connected to interact with any suitable environment, thus setting up a testbed for the reinforcement learning experiments. In this way, testing of RL applications is done according to the plug and play concept. This approach allows consistent, comparable and reproducible results while developing and testing of the RL applications. The toolkit is distributed as a Python package $gym$ \cite{Gym}.

For engineers a challenge is to apply computer science research and development algorithms (e.g. coded in Python) successfully when tackling issues using their models in an engineering-specific environment or modeling language (e.g. Modelica) \cite{MyIEPS2014}, \cite{vanfretti2013unambiguous}.

To ensure a Modelica model's independence of a simulation tool, the Functional Mock-up Interface (FMI) is used. FMI is a tool-independent standard that is made for exchange and co-simulation of dynamic systems' models. Objects that are created according to the FMI standard to exchange Modelica models are called Functional Mock-up Units (FMUs). FMU allows simulation of environment internal logic modelled using Modelica by connecting it to Python using PyFMI library \cite{andersson2016pyfmi}. PyFMI library supports loading and execution of models compliant with the FMI standard.

In \cite{dymrl}, the author declared an aim to develop a universal connector of Modelica models to OpenAI Gym and started implementation. Unfortunately, the attempt of the model integration did not extend beyond a single model simulated in Dymola \cite{Dymola}, which is proprietary software. Also, the connector had other limitations, e.g. the ability to use only a single input to a model in the proposed implementation, the inability to configure reward policy. However, the need for such a tool is well motivated by the interest of the engineering community to \cite{dymrl}. Another attempt to extend this project by Richter \cite{fmirl} did not overcome the aforementioned limitations. In particular, the aim of a universal model integration was not achieved, and a connection between the Gym toolbox and PyFMI library was still missing in the pipeline presented in Figure \ref{fig:teaser}.

Thus, this paper presents ModelicaGym toolbox that serves as a connector between OpenAI Gym toolkit and Modelica model through FMI standard \cite{FMI_standard}.

\subsection{Paper Objective}
Considering a potential to be widely used by both RL algorithm developers and engineers who exploit Modelica models, the paper objective is to present the ModelicaGym toolbox that was implemented in Python to facilitate fast and convenient development of RL algorithms to connect with Modelica models filling the gap in the pipeline (Figure \ref{fig:teaser}).

$ModelicaGym$ toolbox provides the following advantages:
\begin{itemize}
    \item Modularity and extensibility - easy integration of new models minimizing coding that supports the integration. This ability that is common for all FMU-based environments is available out of the box.
    \item Possibility of integration of FMUs compiled both in proprietary (Dymola) and open source (JModelica.org \cite{Jmod}) tools.
    \item Possibility to develop RL applications for solutions of real-world problems by users who are unfamiliar with Modelica or FMUs.
    \item Possibility to use a model of both - single and multiple inputs and outputs.
    \item Easy integration of a custom reward policy into the implementation of a new environment. Simple positive/negative rewarding is available out of the box.
\end{itemize}

\section{Software description}

This section aims to describe the presented toolbox. In the following subsections, toolbox and inheritance structure of the toolbox are discussed.

\subsection{Toolbox Structure}
\begin{figure}[h]
  \centering
  \includegraphics[width=0.7\linewidth]{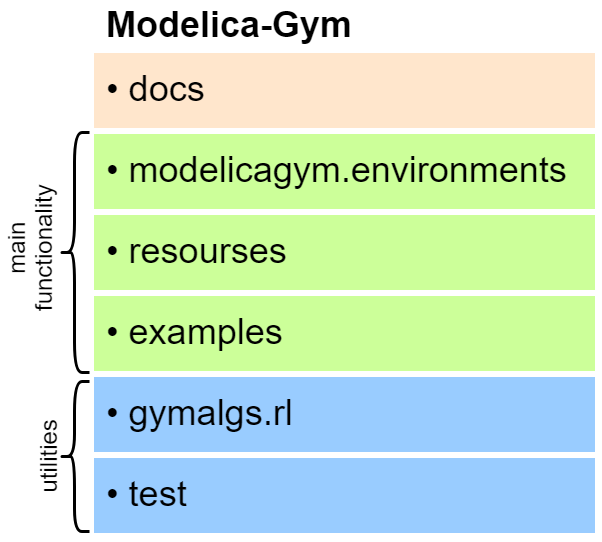}
  \caption{ModelicaGym toolbox structure}
  \label{fig:project-structure-img}
  \Description{Class hierarchy of $modelicagym/environments$}
\end{figure}

ModelicaGym toolbox, that was developed and released in Github  \cite{modelicagym}, is organized according to the following hierarchy of folders (see Figure \ref{fig:project-structure-img}):
\begin{itemize}
    \item $docs$ - a folder with environment setup instructions and an FMU integration tutorial.
    \item $modelicagym/environment$ - a package for integration of FMU as an environment to OpenAI Gym.
    \item $resourses$ - a folder with FMU model description file (.mo) and compiled FMU for testing and reproducing purposes.
    \item $examples$ - a package with examples of:
    \begin{itemize}
        \item custom  environment creation for the given use case (see the next section);
        \item Q-learning agent training in this environment;
        \item scripts for running various experiments in a custom environment.
    \end{itemize}
    \item $gymalgs/rl$ -  a package for Reinforcement Learning algo\-rithms that are compatible with OpenAI Gym environments
    \item $test$ - a package with a test for working environment setup. It allows testing environment prerequisites before working with the toolbox.
\end{itemize}

To create a custom environment for a considered FMU simulating particular model, one has to create an environment class. This class should be inherited from $JModCSEnv$ or $DymolaCSEnv$ class, depending on what tool was used to export a model. More details are given in the next subsection. 

\subsection{Inheritance Structure}\label{sec:struct}

\begin{figure}[h]
  \centering
  \includegraphics[width=0.7\linewidth]{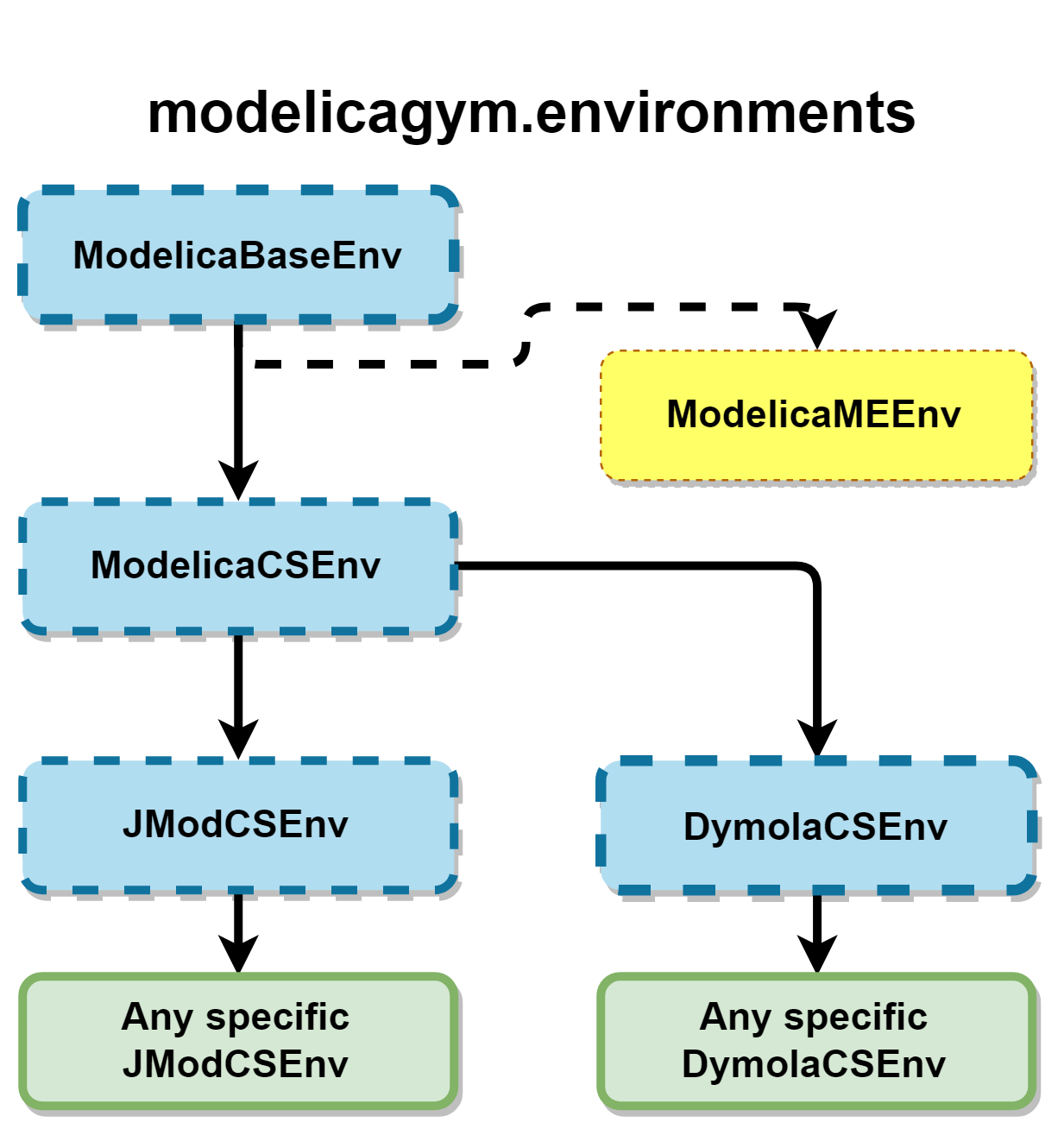}
  \caption{Class hierarchy of the $modelicagym/environments$}
  \label{fig:class-hierarchy-img}
  \Description{Class hierarchy of the $modelicagym/environments$}
\end{figure}

This section aims to introduce a hierarchy of \textit{modelicagym/environ\-ments} that a user of the toolbox needs to be familiar to begin exploitation of ModelicaGym toolbox for his purpose.
The inheritance structure of the main classes of the toolbox is shown in Figure \ref{fig:class-hierarchy-img}.

Folder $modelicagym/environments$ contains the implementation of the logic that is common for all environments based on an FMU simulation. Main class $ModelicaBaseEnv$ is inherited from the $gym.Env$ class (see  Figure \ref{fig:class-hierarchy-img}) to declare OpenAI Gym API. It also determines internal logic required to ensure proper functioning of the logic common for all FMU-based environments.

$ModelicaBaseEnv$ class is inherited by \textit{ModelicaCSEnv} and  \textit{ModelicaMEEnv}. These abstract wrapper classes are created for structuring logic that is specific to FMU export mode: co-simulation or model-exchange respectively. Note, that model-exchange mode is currently not supported.

Two classes $JModCSEnv$ and $DymolaCSEnv$ that inherit \textit{ModelicaCSEnv} class are created to support an FMU that is compiled using Dymola and JModelica respectively (refer to  Figure \ref{fig:class-hierarchy-img}). Any specific implementation of an environment integrating an FMU should be inherited from one of these classes. Further in this section, details of both OpenAI Gym and internal API implementation are discussed.

$ModelicaBaseEnv$ declares the following $Gym$ API:
\begin{itemize}
    \item $reset()$ - restarts environment, sets it ready for a new experiment. In the context of an FMU, this means setting initial conditions and model parameter values and initializing the FMU, for a new simulation run.
    \item $step\ (action)$ - performs an $action$ that is passed as a parameter in the environment. This function returns a new state of the environment, a reward for an agent and a boolean flag if an experiment is finished. In the context of an FMU, it sets model inputs equal to the given action and runs a simulation of the considered time interval. For reward computing $\_reward\_policy()$ internal method is used. To determine if experiment has ended $\_is\_done()$ internal method is used.
    \item $action\_space$ - an attribute that defines space of the actions for the environment. 
    It is initialized by an abstract method $\_get\_action\_space()$,
    that is model-specific and thus should be implemented in a subclass.
    \item $observation\_space$ - an attribute that defines state space of the environment. 
    It is initialized by an abstract method $\_get\_observation\_space()$, 
    that is model specific and thus should be implemented in a subclass.
    \item $metadata$ - a dictionary with metadata used by $gym$ package.
    \item $render()$ - an abstract method, should be implemented in a subclass. It defines a procedure of visualization of the environment's current state.
    \item $close()$ - an abstract method, should be implemented in a subclass. It determines the procedure of a proper environment shut down.
\end{itemize} 

To implement the aforementioned methods, a configuration attribute with model-specific information is utilized by the $Modelica\-Base\-Env$ class. 
This configuration should be passed from a child-class constructor to create a correctly functioning instance. This way, using the model-specific configuration, model-independent general functionality is executed in the primary class.
The following model-specific configuration is required:

\begin{itemize}
    \item $model\_input\_names$ - one or several variables that represent an action performed in the environment.
    \item $model\_output\_names$ - one or several variables that represent an environment's state.
    
\textbf{Note:} Any variable in the model (i.e. a variable that is not defined as a parameter or a constant in Modelica) can be used as the state variable of the environment. On the contrary, for proper functionality, only model inputs can be used as environment action variables. 

    \item $model\_parameters$ - a dictionary that stores model parameters with the corresponding values, and model initial conditions.
    \item $time\_step$ - defines time difference between simulation steps.
    \item \textit{(optional)} $positive\_reward$ - a positive reward for a default reward policy. It is returned when an experiment episode goes on.
    \item \textit{(optional)}  $negative\_reward$ - a negative reward for a default reward policy. It is returned when an experiment episode is ended.
\end{itemize}

However, $ModelicaBaseEnv$ class is defined as abstract, because some internal model-specific methods have to be implemented in a subclass (see \ref{fig:class-hierarchy-img}). The internal logic of the toolbox requires an implementation of the following model-specific methods:
\begin{itemize}
    \item \textit{\_get\_action\_space(), \_get\_observation\_space()} - describe variable spaces of model inputs (environment action space) and outputs (environment state space), using one or several classes from $spaces$ package of OpenAI Gym. 
    \item $\_is\_done()$ - returns a boolean flag if the current state of the environment indicates that episode of an experiment has ended. It is used to determine when a new episode of an experiment should be started.
    \item \textit{(optional)} $\_reward\_policy()$ - the default reward policy is ready to be used out of box. The available method rewards a reinforcement learning agent for each step of an experiment and penalizes when the experiment is done. In this way, the agent is encouraged to make the experiment last as long as possible. However, to use a more sophisticated rewarding strategy, $\_reward\_policy()$ method has to be overridden.
\end{itemize}

Examples and experiments will be discussed in the next section.

\section{Use Case: Cart-Pole Problem}
In this section, a use case of the toolbox set up and exploitation is presented. For this purpose, a classic Cart-Pole problem was chosen.

\subsection{Problem Formulation} \label{sec:problem}
\begin{figure}
  \centering
  \includegraphics[width=\linewidth]{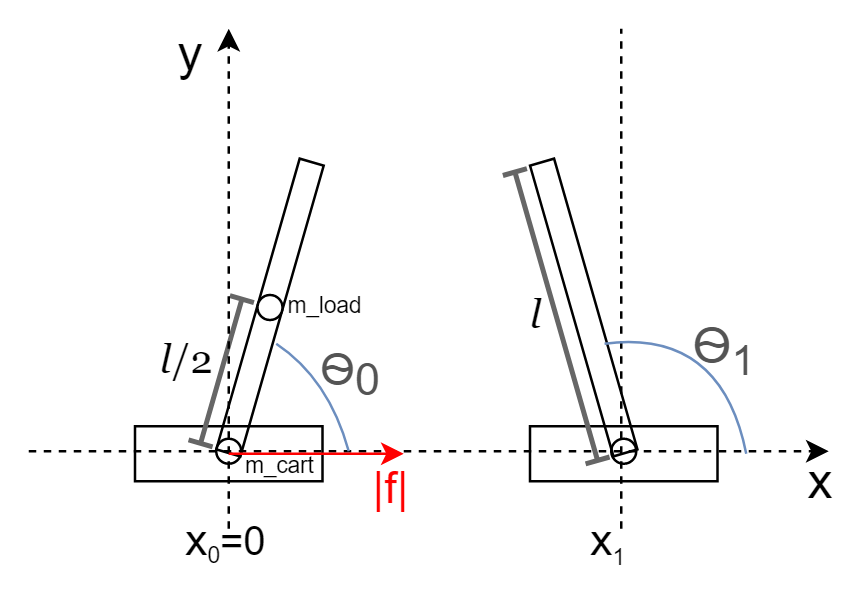}
  \caption{Cart-Pole system}
  \label{fig:cp}
  \Description{Cart-Pole system}
\end{figure}

The two-dimensional Cart-Pole system includes a cart of mass $m\_cart$ moving on a 1-d frictionless track with a pole of mass $m\_pole$ and length $l$ standing on it (see Figure \ref{fig:cp}). 
Pole's end is connected to the cart with a pivot so that the pole can rotate around this pivot. 

The goal of the control is to keep the pole standing while moving the cart. At each time step, a certain force $f$ is applied to move the cart (refer to Figure  \ref{fig:cp}). In this context, the pole is considered to be in a standing position when deflection is not more than a chosen threshold. Specifically, the pole is considered standing if at $i$-th step two conditions $|\theta_i-90^{\circ}|\leq \theta_{threshold}$ and $|x_i|\leq x_{threshold}$ are fulfilled. Therefore, a control strategy for standing upright in an unstable equilibrium point should be developed. It should be noted, that an episode length serves as the agent's target metric that defines how many steps an RL agent can balance the pole.

In this particular case, a simplified version of the problem was considered meaning that at each time step force magnitude is constant, only direction is variable. In this case the constraints for the system are a) moving cart is not further than $2.4$ meters from the starting point, $x_{threshold}=2.4 m$; b) pole's deflection from the vertical is not more than $12$ degrees is allowed, i.e. $\theta_{threshold}=12^{\circ}$.

\subsection{Modelica Model}

\begin{figure*}[h]
  \centering
  \includegraphics[width=\textwidth]{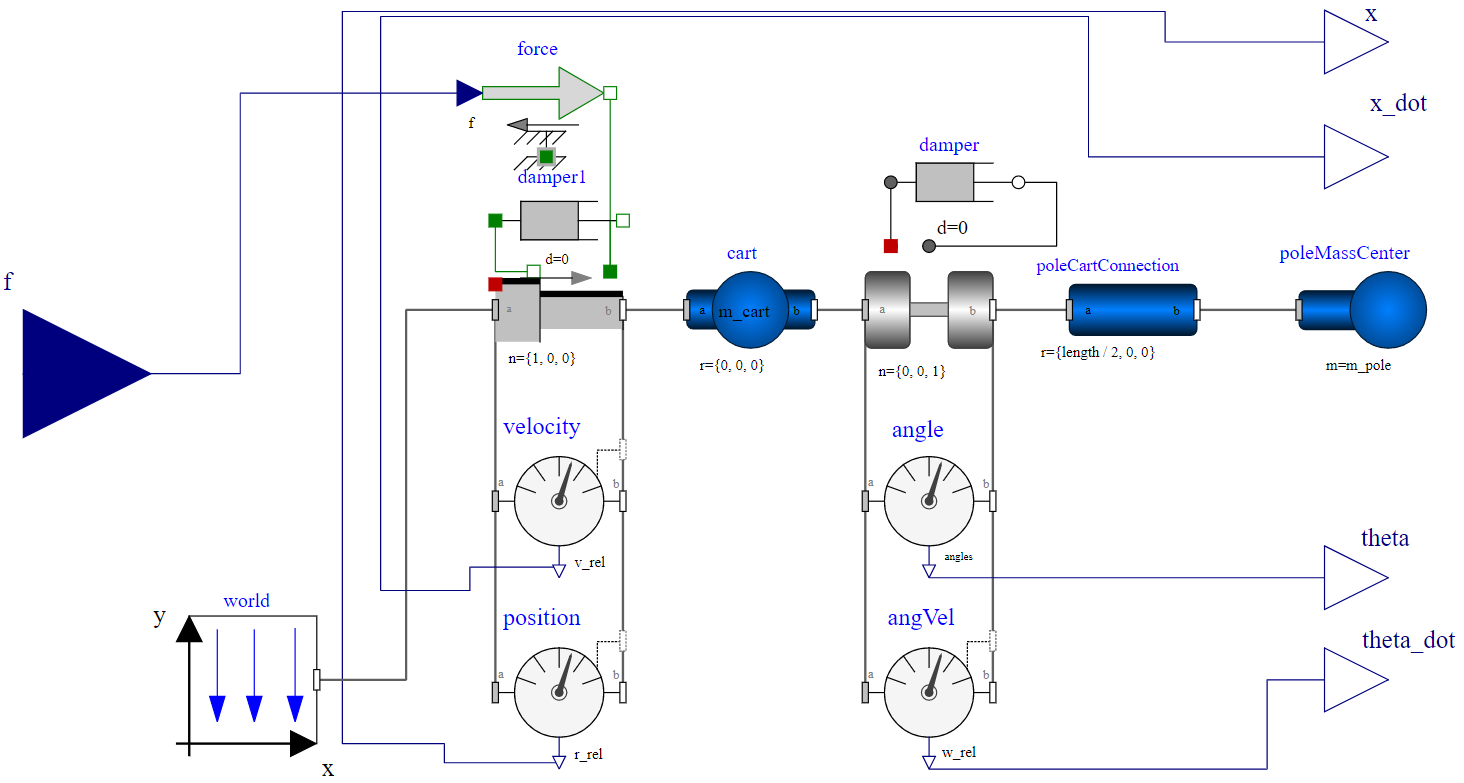}
  \caption{Cart-Pole model. Modelica model structure in OpenModelica \cite{fritzson2006openmodelica}.}
  \label{fig:model-diagram}
  \Description{Cart-Pole model. Modelica model structure in OpenModelica \cite{fritzson2006openmodelica}}
\end{figure*}

A convenient way to model the Cart-Pole system is to model its parts in the form of differential and algebraic equations and to connected the parts together (refer to Figure \ref{fig:model-diagram}). In addition, the elements of the Cart-Pole system can be instantiated from the Modelica standard library. This facilitates the modeling process. However, several changes to the instances are required.

 Thus, to use $Modelica.Mechanics.MultiBody$ efficiently, the modeling problem was reformulated. The pole can be modeled as an inverted pendulum standing on a moving cart. Center of pole's mass is located at an inverted pendulum's bob. To model the pole using the standard model of a pendulum, the following properties have to be considered: a) the length of the pendulum is equal to half of the pole's length; b) a mass of the bob is equal to the mass of the pole. Pendulum's pivot is placed in the centre of the cart. It can be assumed that the mass of the cart can be concentrated in this point as well. Also, a force of the magnitude $|f|$ is applied to the cart to move the pivot along the 1-d track.

As a result, using the standard pendulum model $Modelica.Me\-cha\-nics.MultiBody.Examples.Elementary$, the example in \cite{dymrl} and elements from the Modelica standard library, the model was composed. In contrast to the model in \cite{dymrl}, the developed model is structurally simpler, and its parameters are intuitive. 
To simulate the developed model, an FMU was generated in JModelica.org (see Figure \ref{fig:model-diagram}).

\subsection{Cart-Pole FMU Integration}
To integrate the required FMU using ModelicaGym toolbox, one should create an environment class inherited according to the inheritance structure presented in Section \ref{sec:struct}. To this end, the model's configuration should be passed to the parent class constructor. Furthermore, some methods that introduce model-specific logic should be implemented. In this section, these steps to integrate a custom FMU are discussed.\footnote{One can find a detailed tutorial in the toolbox documentation \cite{modelicagym}. It describes the implementation in a step-wise manner with detailed explanations. This tutorial allows toolbox users to get started quickly.}

To start, an exact FMU specification, which is determined by a model, should be passed to the primary parent class constructor as a configuration. This way, the logic that is common to all FMU-based environments is correctly functioning. 

Therefore, the Modelica model's configuration for the considered use case is given below with explanations.

Initial conditions and model parameters' values are set automa\-ti\-cally when the environment is created. For the considered model these are:
\begin{itemize}
    \item $theta\_0$ - initial angle value of the pole (in rad). This angle is measured between the pole and the positive direction of X-axis (see Figure \ref{fig:cp}).
    \item $theta\_dot\_0$ - initial angular velocity of a pole (in rad/s);
    \item $m\_cart$ - a mass of a cart (in kg);
    \item $m\_pole$ - a mass of a pole (in kg).
\end{itemize}

Environment state is represented by the Modelica model outputs that are generated at each simulation step. For the considered model the state variables are:
\begin{itemize}
    \item $x$ - a cart position (in m);
    \item $x\_dot$ - a cart velocity (in m/s);
    \item $theta$ - the pole's angle (in rad) that is initialized with $theta\_0$;
    \item $theta\_dot$ - the angular velocity of pole (in rad/s) which is initialized with $theta\_dot\_0$
\end{itemize}

The action is presented by the magnitude of the force $f$  applied to the cart at each time step. According to the problem statement (see Section \ref{sec:problem}), the magnitude of the force is constant and chosen when the environment is created, while the direction of the force is variable and chosen by the RL agent.


Listing \ref{lst:CP} gives an example of the configuration for the Cart-Pole environment that has to be passed to the parent class constructor (\textit{ModelicaBaseEnv} in Figure \ref{fig:class-hierarchy-img}).

\begin{lstlisting}[language=Python, caption=Environment configuration for the Cart-Pole exam\-ple, label={lst:CP}]
config = {
'model_input_names': 'f',
'model_output_names': ['x', 
                       'x_dot', 
                       'theta', 
                       'theta_dot'],
'model_parameters':{'m_cart': 10, 
                    'm_pole': 1,
                    'theta_0':85/180*math.pi, 
                    'theta_dot_0': 0},
'time_step': 0.05,
'positive_reward': 1, 
'negative_reward': -100
}
\end{lstlisting}

For the Cart-Pole example, a specific $JModelicaCSCartPoleEnv$ class that relates the settings of the Cart-Pole environment and general functionality of the toolbox was created. 
This class is inherited from the $JModCSEnv$ class according to the toolbox inheritance structure requirements (see Figure \ref{fig:class-hierarchy-img}). 
The \textit{JModelicaCSCartPoleEnv} class was written such that all Modelica model parameters were made class attributes. So that one can configure the environment during an experiment setup. This eases running experiments using the created environment.

 To finish the integration, several model-specific methods were implemented and are briefly discussed below:

\begin{itemize}
    \item $\_is\_done()$ checks if the cart position and pole's angle are inside of the required bounds that are defined by thresholds. This method returns $False$ if the cart is not further than $2.4$ meters from the starting point and pole deflection from vertical position is less than 12 degrees. Otherwise, returns $True$, as the pole is considered as fallen, therefore, the experiment episode is ended.
    \item $\_get\_action\_space()$ returns a \textit{gym.spaces.Discrete} action space of size 2, because only 2 actions \textit{push left} and \textit{push  right} are available for the agent.
    \item $\_get\_observation\_space()$ returns a \textit{gym.spaces.Box} state space with specified lower and upper bounds for continuous state variables.
    \item $render()$ visualizes Cart-Pole environment in the current state, using built-in $gym$ tools.
    \item $step\ (action)$ method was overridden to implement expected action execution, i.e. fixed force magnitude, but variable direction at each experiment step. A sign of the force determines the direction: positive - push the cart to the right, otherwise - push to the left. $close()$ also was overridden to allow a proper shut down of a custom rendering procedure.
\end{itemize}

This way, a custom FMU that is exported in co-simulation mode using JModelica.org and simulates the Cart-Pole environment was integrated to Gym in a couple of straightforward steps. The configured environment allows running experiments in the plug and play manner thanks to the utilization of \textit{ModelicaGym} toolbox.

\section{The Cart Pole Simulation Set Up}

This section aims to explain the components of the Cart Pole experiment set up that could serve as an example for the ModelicaGym user when setting up another experiment. 
\subsection{Experiment Procedure}\label{sec:exp}

To verify the correctness of the FMU integration, several experiments were performed in the implemented Cart-Pole environment according to the pipeline in Algorithm \ref{alg:exp}. For each experiment, a set of parameters that define the Cart-Pole environment was created and set as an input of the procedure. The number of episodes of Q-learning agent's training in reinforcement learning is the input parameter in Algorithm \ref{alg:exp} as well. The value of \textit{n\_episodes} parameter was of the same value for all the experiments in order to maintain equal conditions for further comparison of the simulation results. To obtain a  statistically representative sample, the training was repeated in the restarted environment. 
The output that includes episodes' lengths and total execution time was saved to ease further results analysis and visualization.

\begin{algorithm}[h]
\SetAlgoLined
\SetKwInOut{Parameter}{Parameters}
\Parameter{$n\_repeats$ - number of experiment repeats to perform, $n\_episodes$ - number of episodes to perform in one experiment, $env\_config$ - parameters required to configure the environment}
\KwResult{lists of length $n\_repeats$ with experiment execution times, matrix of shape $(n\_repeats, n\_episodes)$ with episodes' lengths}
\BlankLine
create $env$ with $env\_config$\;
\For{$i=1$ \KwTo $n\_repeats$}{
train Q-learning agent in $env$ \;
append episodes' lengths and execution time to result\;
reset $env$\;
}
\caption{Experiment procedure}
\label{alg:exp}
\end{algorithm}

Following the established procedure (see Algorithm \ref{alg:exp}), four experiments on varying input parameters that influence the outcome of reinforcement learning were established. These experiments are 1) variation of force magnitude, 2) variation of a cart-pole mass ratio, 3) variation of a positive-negative reward ratio, 4) variation of a simulation time step. The values of the changed parameters are given for each experiment in Table \ref{tab:exp-summary}.

\subsection{Q-learning Agent}

In Section \ref{sec:exp}, a Q-learning agent was mentioned in the context of the input parameter settings in Algorithm \ref{alg:exp}. This section aims to explain in detail the role and the set up of the Q-learning agent in Q-learning algorithm for the Cart-Pole experiment using Mo\-de\-licaGym toolbox.
In ModelicaGym toolbox the Q-learning agent is implemented using $QLearner$ class from \textit{gymalgs/rl} package of the toolbox. 

In general Q-learning algorithm assumes discrete state and action spaces of an environment. Therefore, a continuous state space of the Cart-pole environment was discretized by splitting an interval of possible values for each state variable in 10 bins. To take into account possible values outside the interval, the most left bin is bounded with $-\inf$ from the left, while the most right bin is bounded with $+\inf$ from the right. These bins are utilized for encoding the current environment state and getting an index of the state in a Q-table. The Q-table is populated by concatenating indexes of the bins where each of four state variables belongs to. Moreover, the Q-table that represents agents' belief about an optimal control policy is initialized randomly with values uniformly distributed in the interval $[-1;1]$.

According to the problem formulation (see Section \ref{sec:problem}), the intervals for the variables were chosen as  follows:
\begin{itemize}
    \item Cart's position $x$ - $[-2.4;2.4]$.
    \item Cart's velocity $x\_dot$ - $[-1;1]$.
    \item Pole's angle $theta$ - $[\frac{(90-12)}{180}\pi;\frac{(90+12)}{180}\pi]$.
    \item Pole's angle velocity $theta\_dot$ - $[-2;2]$.
\end{itemize}

The Q-learning algorithm is parametrized not only by external parameters that are defined by the environment, but also by its intrinsic parameters. Thus, the following intrinsic parameters of the algorithm and their values were utilized:
\begin{itemize}
    \item $learning\_rate=0.2$ -  a part of Q-value that is updated with each new observation. The chosen value makes agent replace only 20\% of the previous knowledge.
    \item $discount\_factor=1$ -  defines the importance of future reward. The chosen value encourages infinitely long runs.
    \item $exploration\_rate=0.5$ - determines exploration (random action choice) probability at each step. In this case, agent chooses random action with probability equal to 0.5 at the first step.
    \item $exploration\_decay\_rate=0.99$ - means an exploration probability decay at each step. Slow decay was chosen to let the agent explore more in the early learning phase, while to exploit learned policy after significant training time.
\end{itemize}
 
The Q-learning training procedure that is utilized in the experiment procedure in Algorithm \ref{alg:exp} was carried out according to Algorithm \ref{alg:qlern}. 

The Q-learning algorithm uses a concept of Q-value - a proxy that determines the utility of an action in a state. A set of Q-values, which are assigned to all state-action pairs, forms a Q-table. It represents the agent's belief about the optimal control policy.
In the training process, Q-learning agent uses information about the current environment state, action to be performed and the resulting state after the action is applied. The latter updates the Q-table. The update of the Q-values is mathematically formulated as follows:
\begin{align}
     Q(s,a) \leftarrow Q(s,a) + \alpha [r(s,a) + \gamma \cdot max_a Q(s',a) - Q(s,a)],
     \label{eq:upd}
\end{align}
 where $\alpha$ is a learning rate, $ \gamma$ is a discount factor, $s$ - a starting state, $s'$ - a resulting state, $a$ - an action that led from $s$ to $s'$, $r(s, a)$ - a reward received by the agent in the state $s$ after performing the action $s'$, $Q(s,a),\ Q(s',a)$ - the Q-values that are defined by the starting state and the action or the resulting state and the action correspondingly.

To solve the exploration-exploitation trade-off in the training procedure, the Q-learning agent utilizes an \textbf{$\epsilon$-greedy policy}. 
The policy name originates from parameter $\epsilon$ that is referenced as \textit{explo\-ration\_rate} in Algorithm \ref{alg:qlern}. According to the policy, the probability to choose the optimal action is set to $1-\epsilon$. This choice defines exploitation of the best action among already explored actions by the agent. Therefore, to support exploration of the other actions, the next action is chosen randomly with a probability of $\epsilon$.
Adaptive change of parameter $\epsilon$, which is introduced by utilization of \textit{exploration\_decay\_rate}, allows an agent to be more flexible in the early stages of exploration and more conservative in the mature phase.

\begin{algorithm}[h]
\SetAlgoLined
\SetKwInOut{Parameter}{Parameters}
\Parameter{$env$ - environment, 

$max\_number\_of\_steps$ - maximum number of steps allowed in an episode, 

$n\_episodes$ - number of episodes for agent training, 

$visualize$ - a boolean flag if $env$ should be rendered at each step;  

Q-learning parameters: $learning\_rate$, $discount\_factor$, $exploration\_rate$, $exploration\_decay\_rate$ }
\KwResult{A trained Q-learner, 

a list of size $n\_episodes$ with episodes' lengths, 

training execution time in seconds}

\BlankLine
start $timer$;

$ep\_ls \leftarrow []$;

$bins \leftarrow discretize\ state\ space$;

initialize Q-learner with Q-table, given parameters;

\For{$episode=1$ \KwTo $n\_episodes$}{
    $initial\_state \leftarrow env.reset()$\;
    encode $initial\_state$ as a sequence of discretization bin's index\;
    choose initial $action$ randomly\;
    \For{$step=1$ \KwTo $n\_episodes$}{
        \If{$visualize$}{
            render $env$\;
        }
        $state, reward, done \leftarrow env.step(action)$\;
        $enc\_state \leftarrow encode(state, bins)$\;
        update Q-table using $enc\_state$ and $reward$\;
        $action \leftarrow $ choose using Q-table and $\epsilon$-greedy policy\;
        
        \If{$done$ \textbf{OR} step == $max\_number\_of\_steps$}{
            $ep\_ls.append(step)$\;
        }
    }

}

end $timer$;

\Return Q-learner, $ep\_ls$, execution time from $timer$;
\caption{Training a Q-learning agent in an FMU-based environment}
\label{alg:qlern}
\end{algorithm}

\section{Results \& Discussion}

In this section results of experiments that differ in entry Cart Pole parameter values are presented and summarized in Table \ref{tab:exp-summary}.
\subsection{Selection of force magnitude}

According to the formulation of the Cart Pole problem in Section \ref{sec:problem}, one of the variables that influence the experiment is the force that is applied to the cart. Therefore, in this subsection dependency of the learning process of Q-learning algorithm on three different force magnitudes is studied.
These values are chosen with respect to the reference force magnitude, which is the force required to give a cart an acceleration of $1\  m/s^2$. 

Thus, to investigate the Cart Pole system's behaviour, three values of force magnitude were considered: 
\begin{enumerate}
    \item Force magnitude that is significantly smaller than the reference;
    \item Force magnitude that is slightly bigger than the reference;
    \item Force magnitude that is significantly bigger than the reference.
\end{enumerate}
The exact values can be found in Table \ref{tab:exp-summary}. 

\begin{figure}[h]
  \centering
  \includegraphics[width=\linewidth]{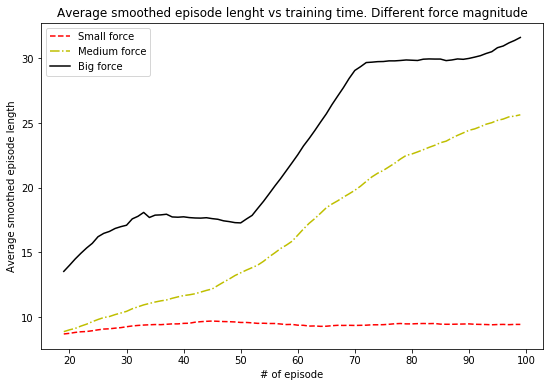}
  \caption{Average smoothed episode length for the force variation experiment.}
  \label{fig:exp-force}
  \Description{For the force bigger than the cart mass agent is learning, while for small force magnitude, it fails to learn.}
\end{figure}

Five experiments were run for each selected force magnitude value with the number of episodes for agent training equal 100. Episodes' lengths were smoothed with a moving average window of size equal to 20 for visualization purpose. In this way, the average smoothed episode length represents the trend. Results are shown in Figure \ref{fig:exp-force}.

\begin{table*}[h]
  \caption{Experiments summary: changed parameter values, average execution time per simulation step}
  \label{tab:exp-summary}
  \begin{tabular}{|c|c|c|c|c|c|c|c|}
    \hline
    \multicolumn{2}{|c|}{Force magnitude} & \multicolumn{2}{|c|}{Cart-pole masses} & \multicolumn{2}{|c|}{Positive-negative reward} & \multicolumn{2}{|c|}{Time step}\\
    \hline
    $|f|,N$ & \shortstack{Seconds per \\simulation step} & $m\_cart; m\_pole, kg$ & \shortstack{Seconds per \\simulation step} & \shortstack{$positive\_reward;$ \\ $negative\_reward$} & \shortstack{Seconds per \\simulation step} & $time\_step, s$ & \shortstack{Seconds per \\simulation step}\\
    \hline
    5 & 0.118 & 1; 10 & 0.111 & 1; -200 & 0.114 & 0.01 & 0.11\\
    \hline
    11 & 0.113 & 5; 10 & 0.112 & 1; -100 & 0.113 & 0.05 & 0.113\\
    \hline
    17 & 0.112 & 10; 10 & 0.118 & 1; -50 & 0.113 & 0.1 & 0.117\\
    \hline
    - & - & 10; 5 & 0.112 & - & - & 0.5 & 0.145\\
    \hline
    - & - & 10; 1 & 0.113 & - & - & 1 & 0.218\\
    \hline
    \multicolumn{8}{|l|}{Average execution time per simulation step, s: 0.122} \\
    \hline
    \multicolumn{8}{|l|}{Average execution time per simulation step, excluding 0.5s, 1s time step experiments, s: 0.114} \\
  \bottomrule
\end{tabular}
\end{table*}

In Figure \ref{fig:exp-force}, as expected, the episode length growth was observed for the moderate and big magnitude of the applied force. In this problem the episode's length is a reinforcement learning agent's target metric. Therefore, it can be stated that the agent is learning in these cases. Moreover, with bigger force magnitude higher value of average episode length can be reached, meaning the agent is learning faster.

However, in the third case, the agent fails to learn. In Figure \ref{fig:exp-force} we observe a plateau that is close to the initial level in this case. The reason is that with such a small magnitude of the applied force to the cart, it is not possible to cause enough influence on the system to balance the poll within the given constraints.

\subsection{Selection of Cart-pole mass ratio}
In the Cart Pole problem another physical parameter that influences the control and, therefore, the reinforcement learning process, is cart-pole mass ratio.

Thus, to observe this influence, the system's behaviour where five pairs of the cart and pole masses with different ratio were considered, is studied in this section.

These five pairs of the mass ratio are selected as follows:
\begin{enumerate}
    \item The pole's mass is significantly bigger than the mass of a cart.
    \item The pole's mass is two times bigger than the mass of a cart.
    \item The pole's mass is equal to the mass of a cart.
    \item The pole's mass is two times smaller than the mass of a cart.
    \item The pole's mass is significantly smaller than the mass of a cart.
\end{enumerate}

Exact values of the selected mass ratio are shown in Table \ref{tab:exp-summary}. In Table \ref{tab:exp-summary} for each experiment with the selected mass ratio, the number of episodes for agent training is equal to 200.

The observed system's behaviour in most scenarios of the selected Cart-pole mass ratio indicates that the agent is able to learn, showing a good performance, regardless of what is heavier: a cart or a pole. This was observed in 4 out of 5 cases when the RL agent's ability to perform the required task increased with an augment in the training time.\footnote{One can find the results and visualizations in the toolbox documentation \cite{modelicagym}} In the mentioned four cases, the agent reached the same level of performance that is measured by a smoothed average episode length of around 40 steps. Therefore, it can be concluded that in these cases the chosen values of the cart and pole masses do not influence training speed.

\begin{figure}[h]
  \centering
  \includegraphics[width=\linewidth]{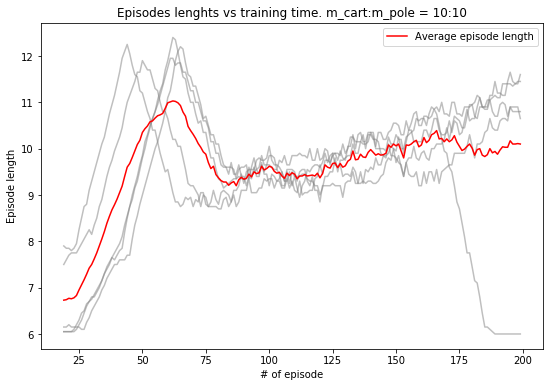}
  \caption{Smoothed episode length for experiment (3)}
  \label{fig:exp-eq-m-ratio}
  \Description{With equal cart and pole masses agent fails to learn.}
\end{figure}

However, for the case (3), when masses of cart and pole are equal,   the observed system's behaviour is extremely different (see Figure \ref{fig:exp-eq-m-ratio}). The difference is that episode length does not increase along with the number of episodes. 

For the visualization purpose, the episode lengths were smoothed with a moving average of window size 20. 
Average smoothed episode length (in red in Figure \ref{fig:exp-eq-m-ratio}) represents the trend of the experiment. 
There is a plateau in epi\-so\-de length at the level of 9-10 steps. This value is almost equal to the initial guess observed at the beginning of the experiment, indicating that the agent fails to train. 

In one of the runs the episode length, which defines the agent's performance, drops considerably to the episode length value of 8 steps.
The reason for such phenomenon may be that the physical system of cart pole of equal masses is very unstable. Therefore, it may be not possible to balance the pole within the given constraints.

\subsection{Selection of Reward ratio}
The aim of this subsection is to observe a dependency of the system behaviour and it's leaning performance on the reward value. Two types of the rewards, positive and negative, are assigned among other input parameters of the experiment. While the positive reward is given when the agent succeeds to hold the pole in a standing position and experiment episode goes on, the negative reward is assigned when pole falls and experiment episode ends. Three different pairs of positive and negative reward values were considered in the experiment (refer to the exact values in Table \ref{tab:exp-summary}):
\begin{enumerate}
    \item The negative reward is so big, that agent has to balance pole for 200 steps to get a non-negative reward.
    \item The negative reward is so big, that agent has to balance pole for 100 steps to get a non-negative reward.
    \item The negative reward is so big, that agent has to balance pole for 50 steps to get a non-negative reward.
\end{enumerate}

\begin{figure}[h]
  \centering
  \includegraphics[width=\linewidth]{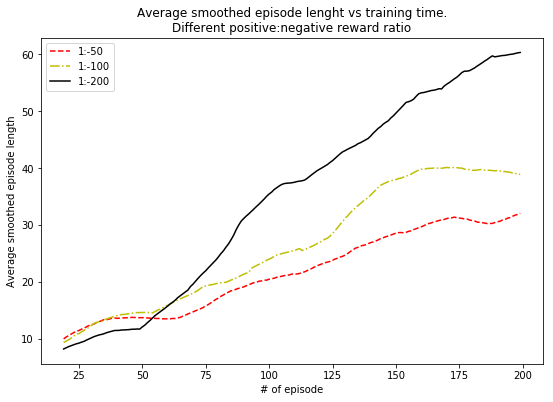}
  \caption{Average smoothed episode length for the reward variation experiment.}
  \label{fig:exp-reward}
  \Description{For bigger negative reward agent learns faster.}
\end{figure}

The length of each experiment has been defined by the number of episodes for training the agent (Figure \ref{fig:exp-reward}).
In order to visualize a trend, the episode's lengths were smoothed with a moving average window of size that is equal to $20$.

According to expected longer time balancing of the pole on the cart, the observed episode's length increases with a training time increase (refer to Figure \ref{fig:exp-reward}).
Besides, it can be noticed that the biggest negative reward's magnitude leads to the best result. In particular, the episode's length increased faster in this case. Also, the final average smoothed episode length, which is an indicator of an agent's capability to solve the balancing problem, get bigger when negative reward's magnitude increases.

On the contrary, when the negative reward's magnitude is smaller, the slower training and even a significant decrease in performance are observed. These decreases could be resolved with more extended experiment duration, but this is not an optimal training result.
The reason may be in the fact that a smaller negative reward's magnitude penalizes bad agent's decisions not strongly enough.

\subsection{Time step variation and Execution time}

Time step defines an interval between two consecutive control actions that are applied to the system. At each iteration the system's behaviour is simulated in this interval. To study the influence of a time step change on the training result, five different values for the time step were considered and presented in Table \ref{tab:exp-summary}. In particular, the smallest simulation time step ($0.01\ s$) appeared to be too small for a real experiment when controlling a Cart-Pole system. While the biggest simulation time step ($1\ s$) is too large for keeping the balance of the system, i.e. the pole will be able to fall within the time step.

It was observed that with a very small time step training is too slow, while it is inefficient for a very big time step (see Table \ref{tab:exp-summary}). When the time step equals to $0.5\ s$, the agent can not overcome the threshold of simulation length that yields four steps. The reason for this is most likely the same as for the simulation time step of $1\ s$. Thus, such learning behaviour is due to the fact that large time steps narrow the ability to control the system for achieving the required goal. In other words, the control action changes too seldom to reach the goal. 

On the other hand, a simulation of the learning process with a big time step takes less execution time than the simulation of the same interval of learning with a smaller time step. This result caused by the fact that additional time is spent for each call of an FMU in each time step, while the FMU simulation time slightly increases.

Thus, to guarantee effective and efficient training, a trade-off between a time step length and execution time of learning has to be found. 
For the considered system, a simulation time step of $0.1\ s$ is a good choice. This is reasonable from the point of view of both training and application.

For all the experiments execution time was measured. Average time per simulation step is summarized in Table \ref{tab:exp-summary}. It was observed, that for the fixed time step, time per simulation step is almost the same for any set of parameters.

\section{Conclusions}
In this project ModelicaGym - the open source universal toolbox for Modelica model integration in OpenAI Gym as an environment to provide more opportunities for a fast and convenient RL application development - is developed.

Thus, both RL solution developers and the engineers using Modelica can benefit from using ModelicaGym that allows for FMU to simulate environment logic, therefore, use reinforcement learning solution for Modelica modelled real-world problems.

Using a classic Cart-Pole control problem example, the ModelicaGym functionality, modularity, extensibility and validity have been presented and discussed. The results of the experiments that were performed on the Cart-Pole environment indicated that integration was successful. 

The toolbox can be easily used by both developers of RL algorithms and engineers who use Modelica models. It provides extensive options for model integration, allows to employ both open source and proprietary Modelica tools. Thus, it is expected to have great value for a broad user audience.

\section{Future Work}
Even though ModelicaGym toolbox is set and ready, there are several extensions that can be made.

First of all, a more sophisticated use case is in focus of our research plans. Power system models having higher modeling complexity are a suitable research object. Especially in the context of applying more advanced reinforcement learning algorithms.

Second, adding reinforcement learning methods that are working out of the box with small customization would enhance the toolbox functionality.

In addition, currently only FMUs exported in co-simulation mode are supported. Thus, one more extension step would be testing and providing functionality for FMUs exported in model-exchange mode. There is such a possibility in the toolbox architecture. However, feedback from the community should be received first to understand a real demand for such functionality.

\begin{acks}
We would like to thank Dr. Luigi Vanfretti from Rensselaer Polytechnic Institute for insightful discussions and for igniting our inspiration to use Modelica.

Authors would like to thank ELEKS (\url{https://eleks.com/}) for funding the Machine Learning Lab at Ukrainian Catholic University and this research.
\end{acks}

\bibliographystyle{ACM-Reference-Format}
\bibliography{sample-base}

\appendix

\section{Setup}
The toolbox was tested on the following environment setup:
\begin{itemize}
    \item Ubuntu 18, 64-bit version
    \item Python 3.6.8
    \item Java 8
    \item Assimulo 2.9
    \item PyFMI 2.3.1
    \item Sundials 2.4.0
    \item Ipopt 3.12.12
\end{itemize}

Listed libraries are required for proper $ModelicaGym$ usage. Modelica tools are required for FMU compilation, so are optional for toolbox usage. If one uses co-simulation FMU exported from Dymola, licence file should be available as well. Authors utilized JModelica 2.4 and Dymola 2017. 

Machine parameters were: Intel-i7 7500U 2.7GHz (3 cores available), 12GB RAM.

\end{document}